\documentclass[12pt,aps,nofootinbib,floats]{revtex4}
\usepackage{graphicx}
\usepackage{epsfig}

\def\simlt{\stackrel{<}{{}_\sim}}
\def\simgt{\stackrel{>}{{}_\sim}}
\begin{document}
\title{Inflationary cosmology in the \\ central region of String/M-theory
moduli space}

\author{R. Brustein\protect\( ^{(1)}\protect \),
S. P. de Alwis\protect\( ^{(2)}\protect \), E. G. Novak
\protect\(^{(1)}\protect \)}

\affiliation{(1) Department of Physics, Ben-Gurion University,
Beer-Sheva 84105, Israel
\\
 (2) Department of Physics, Box 390, University of Colorado,
 Boulder, CO 80309.\\
 \texttt{e-mail:  ramyb@bgumail.bgu.ac.il },
\texttt{dealwis@pizero.colorado.edu},
\texttt{enovak@bgumail.bgu.ac.il}}

\begin{abstract}

The ``central" region of moduli space of M- and string theories is
where the string coupling is about unity and the volume of compact
dimensions is about the string volume.  Here we argue that in this
region the non-perturbative potential which is suggested by membrane
instanton effects has the correct scaling and shape to allow for enough
slow-roll inflation, and to produce the correct amplitude of CMB
anisotropies. Thus, the well known theoretical obstacles for
achieving viable slow-roll inflation in the framework of
perturbative string theory are overcome. Limited knowledge of some
generic properties of the induced potential is sufficient to
determine the simplest type of consistent inflationary model and
its predictions about the spectrum of cosmic microwave background
anisotropies: a red spectrum of scalar perturbations, and
negligible amount of tensor perturbations.

\end{abstract}

\maketitle


Attempts to obtain viable inflationary cosmology in the framework
of string/M-theory have encountered notorious difficulties. In
particular, potentials are typically too steep, so they cannot
provide enough slow-roll inflation\cite{BG,BS,BBSMS}, and models of
``fast-roll" inflation\cite{PBB} have difficulty in reproducing the
observed spectrum of CMB anisotropies\cite{MVDV}.

In \cite{bda2} a scenario for stabilization of string- and
M-theoretic moduli in the central region of moduli space was proposed.
In this scenario, the central region is parameterized by chiral
superfields of D=4, $N=1$ Supergravity (SUGRA), which are all
stabilized at the string scale by stringy non-perturbative (SNP)
effects induced by membrane instantons \cite{bda1}. Supersymmetry
(SUSY) is broken at a lower intermediate scale by field theoretic
effects that shift the stabilized moduli only by a small amount
from their unbroken minima. The cosmological constant can be made
to vanish after SUSY breaking if there is an adjustable constant
of stringy origin.

Here we show that this scenario can accommodate slow-roll inflation
without unnatural tuning of potential parameters, thus overcoming the
aforementioned difficulties. The simplest consistent inflation model is
that of topological inflation \cite{vilenkin,BBSMS}, in which inflating
domain walls are formed at the top of the barrier separating the central
region from the outer (perturbative) region of moduli space, and
subsequently the field roles down to the SUSY preserving minimum in the
central region and inflation ends. We find a model that predicts a
spectral index of scalar perturbations that is less than unity, that has a
negligible amount of tensor perturbations, and that provides an
opportunity for the use of measurements of CMB anisotropies as a probe of
central region moduli dynamics.

The moduli potential is an important ingredient in any string theoretic
cosmological scenario. So any argument which ignores the question of
moduli stabilization fails to address an important cosmological issue and
its conclusions are suspect. The usual mechanisms which generate moduli
potentials in the perturbative regions of moduli space give runaway (or
steep) potentials that vitiate many conclusions coming from models of
brane annihilation, tachyon condensation and Ekpyrosis.  We will discuss
problems with such scenarios and the possible evolution from the outer
perturbative regions to the central region separately\cite{bdn}.  The
point of this paper is that the mechanism that one might expect to exist
in the central region to stabilize the moduli will also generate
sufficient inflation and lead to an acceptable CMB spectrum. So any
additional inflationary mechanism is redundant.



To determine the size of the central region we need to determine
the correct normalization of moduli kinetic terms.  We follow an
argument which was first made by Banks \cite{banks} in the
context of Horava-Witten (HW) theory in the outer region of
moduli space. In the effective 4D theory obtained after
compactification of 10D string theories on a compact volume $V_6$,
moduli kinetic terms are multiplied by compact volume factors
$M_S^8 V_6$ ($M_S$ being the string mass), and $M_{11}^9 V_7$ in
M-theory compactifications ($M_{11}$ being the M-theory scale and
$V_7$  the compact volume\footnote{Note that we are using the
term M-theory in the restricted sense of being that theory whose
low energy limit is 11 dimensional supergravity.}). The curvature
term in the effective 4D action is multiplied by the same volume
factors. We may use the 4D Newton's constant 
\footnote{Note that $m_p$ is the reduced Planck mass $m_p=2.4
\times 10^{18}$ GeV.}
$8\pi G_N=m_p^{-2}$
 to ``calibrate" moduli kinetic terms, and
determine that they are multiplied by factors of the 4D (reduced)
Planck mass squared, $ \Gamma= \int d^4x
\left\{\frac{m_p^2}{2}\sqrt{- g} R +
\frac{m_p^2}{2}\partial\psi\partial\psi\right\}$. This argument
holds for all string/$M$-theory compactifications.

Strictly speaking, the argument we have presented can be expected to be
somewhat modified in the central region, perhaps by different factors of
order unity multiplying the curvature and moduli kinetic terms. But the
effective 4D theory must always take the form of $\Gamma$ with some SUGRA
potential. Using this scaling, we find that the typical distances over
which the moduli can move within field space, while remaining within the
central region, should be a number of order one in units of $m_p$.


We argue that the overall scale of the moduli potential  is
$\Lambda^4=M_S^6/m_p^2$ in compactifications of string theories
and  $\Lambda^4=M_{11}^6/m_p^2$ in compactifications of
$M$-theory. Of course, in the central region we expect that
$M_S\sim M_{11}$.
First, the overall scale of the superpotential is  $M_S^3$ in
string theory compactifications and $M^3_{11}$ for $M$-theory
compactifications. The key point here is that there are no
additional volume factors coming from some or all of the compact
dimensions. This was first noticed in the context of outer region
compactifications of HW\cite{banks}. Banks argued that the
superpotential is generated on the 4D branes, and that since the
superpotential is generated only in 4D theories, it cannot depend
on the compactification volume which is a higher dimensional
object. Note that this scaling argument is very general and
applies to the central region as well. Any additional volume
factors multiplying $M_S^3$ could only result from the existence
of zero-modes associated with the embedding of the wrapped
Euclidean brane in the compact space. However, such zero-modes
will necessarily enhance the number of SUSY's in the effective 4D
theory, and so by non-renormalization theorems a superpotential
could not be generated.  A superpotential of order $M^3_{S/11}$ will in
turn produce a potential of order $\Lambda^4=M_{S/11}^6/m_p^2$.

Our argument is supported by existing explicit calculations of the
non-perturbative superpotential induced by brane
instantons \cite{HM,Moore,Ovrut}. There the induced potential is
calculated by comparing a gravitino correlator in the background
of the brane instanton to an effective theory with instanton
effects integrated out. The comparison shows that the prefactor
contains only factors associated with topological properties of
the compact space.

Let us give two representative examples, type I string theory and
HW theory. We can perform analogous calculations in other
backgrounds, or simply use the duality relations connecting them.
The numerical relations are taken from \cite{bda1}. Consider type
I string theory compactified on a 6D manifold. The string length
is denoted by $l_I$, the string coupling by $g$, and the compact
6D volume by $V=(2 \pi R)^6$. The string scale $M_S$ in type I
theory is related to $l_I$ by $M_S^8=2 (2\pi)^{-7}l_I^{-8}$, the
4D Newton's constant is given by $ G_N=\frac{1}{8} g^2 l_I^8
R^{-6}$ and the Yang-Mills coupling by $ \alpha_{YM}=\frac{1}{2}
g l_I^6 R^{-6}$. These are related by $ G_N/\alpha_{YM}=
\frac{1}{4} g l_I^2 $; therefore the string length is given by
$l_I=2 (g~\alpha_{YM}~8\pi m_p^2)^{-1/2}$, and the string mass by
$ M_S=(4\pi)^{-7/8}(g~\alpha_{YM}~8\pi m_p^2)^{1/2}$. Putting
$m_p=2.4\times 10^{18} GeV$,  we obtain an estimate for the
string mass in type I string theory of $ M_S=2.6\times 10^{17}
GeV \sqrt{\alpha_{YM}/(1/25)}\sqrt {g} $. Since we have defined
$\Lambda^4=M_S^6 m_p^{-2}$, our estimate for the moduli potential
scale $\Lambda$ in type I theory is
\begin{equation}
\Lambda_I=8.6\times 10^{16} GeV
\left(\frac{\alpha_{YM}}{(1/25)}\right)^{3/4} g^{3/4}.
 \label{lamI}
\end{equation}

Now consider compactifications of HW theory. The size of the 11
dimension interval is $\pi \rho$, and the compact 6 volume is
$V=a^6$. The 11D Planck mass $M_{11}$ is related to the 11D
gravitational constant by $ \kappa_{11}^2=M_{11}^{-9} $. The 4D
Newton's constant is given by $G_N= \frac{1}{8\pi^2}
\kappa_{11}^2 a^{-6}\rho^{-1}$ and $ \alpha_{YM}=\frac{1}{2}(4\pi
\kappa_{11}^2)^{2/3} a^{-6}$, so that
$M_{11}=(4\pi)^{1/9}(2\alpha_{YM})^{-1/6} a^{-1}$. Putting in numbers
we find that $ M_{11} \simeq 2\left(\alpha_{YM}/(1/25)\right)^{-1/6}
\frac{1}{M_{GUT} a} M_{GUT} $. Taking $M_{GUT}=3\times 10^{16}$
GeV, $M_{GUT} a\sim .25-1$, allows $M_{11} \pi \rho \sim 1$ which
means that the string coupling is of order unity, $g\sim 1$, and
$M_{11}\simgt M_{GUT}$. So our estimate for the string mass in
central region compactifications of HW theory is $ M_{11}=2.4
\times 10^{17} GeV \frac {1} {4 M_{GUT}a}
\left(\alpha_{YM}/(1/25)\right)^{-1/6}$, and since the potential scales as
$\Lambda^4=M_{11}^6 m_p^{-2}$ we obtain
\begin{equation}
 \Lambda_{HW}\!=\!7.6\!\times\! 10^{16} {\rm GeV}\!
 \left(4 M_{GUT}a\right)^{-3/2}\!
 \left(\frac{\alpha_{YM}}{(1/25)}\right)^{-1/4}.
 \label{lamHW}
\end{equation}
With $\alpha_{YM}\!\sim\!1/25$ at the GUT scale and
$M_{GUT}a\!\sim\!1$ we have $M_{11}\!\sim\!M_S\sim\!10^{17}$GeV,
and $\Lambda_I\!\sim\!\Lambda_{HW}\!\sim\!10^{16}$ GeV.


Let us consider the expected form of the superpotential in the
central region based on the picture discussed in \cite{bda1}. The
theory in each corner of moduli space has non-perturbative
effects that originate from various Euclidean branes wrapping on
cycles of the compact space. However, in the outer region they
will only give runaway potentials that take the theory to the
zero coupling and decompactification limit. On the other hand
every perturbative theory has strong coupling (S-dual) and small
compact volume (T-dual) partners. From the point of view of one
perturbative theory, the potential in the dual theory is trying
to send the original theory to the strong coupling and/or zero
compact volume limit. Thus it would seem that in the universal
effective field theory in the central region  there are competing terms in
the potential.  If the signs of the prefactors are the same and
(as expected) are of similar magnitude, the potential would contain
a minimum at $g\sim O(1)$ and the size of the internal manifold would be of the
order of the string scale.

To illustrate this consider the S-dual type I and Heterotic
SO(32) (I-HO) theories. The S-duality relations between them are
$\phi_{I} = -\phi_{HO}$, $g_{I}=\frac{1}{g_{H}}$,
$l_{I}^{2}=g_{HO}l^{2}_{HO}$. The string coupling is related to
the expectation value of the dilaton by $g=\langle e^{\phi
}\rangle$, and $l_{I,HO}$ refers to the string length in I/HO
theories, respectively. Note that the relation between scales is
defined in terms of the expectation value of the (stabilized)
dilaton.  In this case terms in the superpotential $W$ come from a Euclidean
$D_{5}$ brane wrapping the whole compact six space, or a Euclidean $D_{1}$
wrapping a one cycle in the compact six-space on the type I side,
and a Euclidean $F_{5}$ brane wrapping the whole compact six
space, or a Euclidean $F_{1}$ brane wrapping a one cycle in the
compact six-space on the HO side. The leading order expressions
(up to pre-factors whose size we have estimated to be $M_S^3$)
for these can be read off from table II of
\cite{bda1}.

In the central region we then expect both these competing effects to be
present(we hope to discuss this more concretely in a future publication).
As explained in \cite{bda2}, the minimum should be supersymmetric
$W'=W=0$, where the latter may happen, for instance, if there is an
R-symmetry under which W has an R-charge of 2 \cite{BBSMS}. 

The moduli potential originates from  brane instantons which
depend exponentially on string coupling $g$, or on the size of the
11D interval or circle $\rho/l_{11}$, and on compactification
radii $R/l_s$, as we have just described. Calling them
generically $\psi$, the 4D action takes the form $ \Gamma=\int
d^4 x \left\{\frac{1}{2} m_p^2 \partial\psi
\partial\psi - \Lambda^4 v(\psi)\right\}.
$ The overall scale of the potential was estimated previously as
$\Lambda=M_S^{3/2}m_p^{-1/2}$, and the potential $v(\psi)$ can be
approximated in the central region by  a polynomial function.
Canonically normalizing the moduli in order  to later compare to
models of inflation we define $\phi= m_p\psi$, and obtain
$
\Gamma=\int d^4 x \left\{ \frac{1}{2} \partial\phi \partial\phi -
\Lambda^4 v(\phi/m_p) \right\}.
$

By assumption, the moduli potential has a SUSY preserving minimum
where it vanishes in the central region. Additionally, it has to
vanish at infinity (the extreme outer parts of moduli space),
since there the non-perturbative effects vanish exponentially, and
the 10D/11D SUSY is restored. Since $v(\phi)$ vanishes at the
minimum in the central region and $v(\phi)$ vanishes at infinity,
its derivative $\delta v/\delta\phi$ needs to vanish somewhere in
between. Since the potential is increasing in all directions away
from the minimum in the central region, that additional extremum
needs to be a maximum.

This means that in any direction there is at least one maximum
separating the central region and the outer region. As we
discussed, the distance of this maximum from the minimum is a
number of order one in units of $m_p$. Note that since we are
dealing with multidimensional moduli space the maxima in
different directions do not need to coincide, so in general they
are saddle points, but we focus on a single direction in field
space. The simplest models without additional tuning of the
parameters will not have additional stable local minima with
non-vanishing energy density. This would require tuning of four
coefficients in the superpotential (or Kahler potential).

\begin{figure}[ht]
\begin{center}
\rotatebox{-90}{\epsfig{file=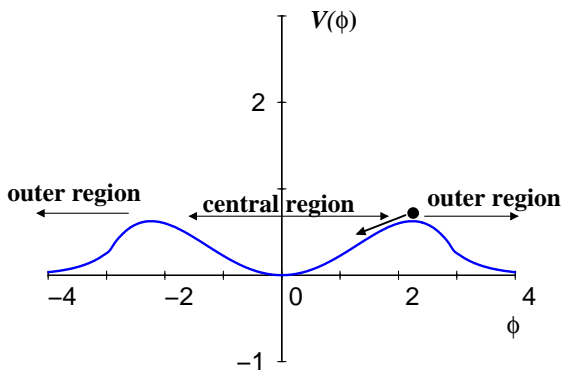,width=5.0cm}}
 \caption{An
example of the expected form of the moduli potential in the central
region of moduli space. Shown are the SUSY preserving minimum at
$\phi=0$ (arbitrary choice), and the maximum near $\phi=2$. The
inflaton rolls from the top of the barrier to the minimum as
indicated. $V(\phi)$ is in units of $M_S^6m_p^{-2}$ and $\phi$ is
in units of $m_p$. } \label{pot}
\end{center}
\end{figure}

The simplest model of central region inflation is that of
topological inflation from the top of the barrier between the
central and outer region. If initially the field is randomly
distributed and can settle down on either side of the barrier,
then typically some inflating domain wall will be produced
\cite{vilenkin} (see also \cite{BBSMS}). Calling the thickness (in
configuration space) of the wall $\delta$, and the distance (in
field space) between the central and outer region minima
$\Delta$, then the balance between gradient and potential
energies requires that
$\left(\frac{\Delta}{\delta}\right)^2\simeq\Lambda^4$. Einstein's
equation relates the Hubble parameter $H$ and $\Lambda$ by
$H^2=\frac{1}{3 m_p^2} \Lambda^4$. According to \cite{vilenkin} an
inflating domain wall will be produced provided that its width is
comparable to the horizon size $H^{-1}$, that is $\delta H \simgt
1$. This requirement is then equivalent to $\Delta\simgt m_p$,
which is clearly satisfied in the case at hand. The previous line
of reasoning shows that generic initial conditions indeed produce
inflating regions. At the end of inflation the inflaton field
settles into its central region SUSY preserving minimum.

We now wish to show that the model we are proposing can produce
enough inflation to resolve the cosmological problems that
inflation is supposed to resolve. The potential near the maximum
of the barrier at $\phi=\phi_{\rm max}$ can be approximated by the
form $\Lambda^4 \left[1- (\frac{\phi-\phi_{max}}{\mu})^2\right]$,
with $\mu^2=\frac{2 m_p^2}{ |v''(\phi_{\rm max})|}$. We recognize
this model as a ``small single field" $p=2$ inflation model
according to the classification of \cite{kinney1}. This model is
similar in many of its phenomenological consequences to
``natural inflation" \cite{natural}.

Recall the definitions of the slow roll parameters ( we use the
conventions of \cite{kolb,kinney1,kinney2} except that we use the
reduced Planck mass $m_P^2=1/(8\pi G_N$)),
$\epsilon=\frac{m_p^2}{2}\left(\frac{V'}{V}\right)^2$,
 and $\eta={m_p^2}\left(\frac{V''}{V}\right)-
\frac{m_p^2}{2}\left(\frac{V'}{V}\right)^2$. The end of inflation
is reached when $\epsilon=1$,  and the number of inflationary e-folds
is given by
\begin{equation}
N(\phi,\phi_{\rm end})=\frac{1}{\sqrt{2}m_p}
\int\limits_{\phi}^{\phi_{\rm end}}
\frac{1}{\sqrt{\epsilon(\phi)}} d\phi .
 \label{efolds}
\end{equation}
For our particular case $\sqrt{\epsilon(\phi)}=
-\sqrt{2}\frac{m_p}{\mu^2}\frac{ \phi-\phi_{\rm max}
}{1-\left(\frac{\phi-\phi_{\rm max}}{\mu}\right)^2}$. The
slow-roll conditions are obeyed if the slow roll parameters are
small, which near the top of the potential they indeed are $\epsilon
\sim 2\frac{m_p^2}{\mu^4}(\phi-\phi_{\rm max})^2$ and $\eta\sim
v''(\phi_{\rm max})$, provided that $v''(\phi_{\rm max})$ is small
enough (see below).

Since the required number of e-folds to resolve the
cosmological problems is about 60, the initial value of the field
needs to be close enough to the maximum of the potential such that
$N(\phi_{\rm initial},\phi_{\rm end})\simgt 60$. Integrating
eq.(\ref{efolds}), assuming that $\phi_{\rm end}\sim \mu$ we
obtain $ \phi_{\rm initial}-\phi_{\rm max}\simeq\mu
e^{-\left[120\left(\frac{m_p}{\mu}\right)^2\right]}$. The
strongest  constraint comes from requiring that quantum
fluctuations are not larger than this estimate. The strength of
quantum fluctuations is $\delta\phi\sim \frac{H}{2\pi}$, which
implies that $\frac{\mu}{m_p}
e^{-\left[120\left(\frac{m_p}{\mu}\right)^2\right]}\simgt
\frac{1}{2\pi}\frac{\Lambda^2}{m_p^2}$, where we have used
$H^2=\frac{1}{3 m_p^2} \Lambda^4$. This constraint can be satisfied
provided that $\left(\frac{m_p}{\mu}\right)^2\simlt 1/10$, that is
provided that $v''(\phi_{\rm max})\simlt 1/6$.

The amplitude of CMB anisotropies can be obtained by standard use
of the  slow roll equations, which lead to
\hbox{$P_{\zeta}^{1/2}\!=\!\frac{15}{2}\frac{\delta\rho}{\rho}\!=\!{\frac{\sqrt
3}{2\pi}} \frac{1}{m_p^3} \frac{V^{3/2}}{V'}\!\simeq\! 10^{-4}$}.
Following \cite{kinney1} we integrate eq.(\ref{efolds}), assuming
\hbox{$\phi_{end}\!\sim\!\mu$} to obtain $\phi_{CMB}\!=\!\mu
e^{-\left[100\left(\frac{m_p}{\mu}\right)^2\right]}$, where
$\phi_{\rm CMB}$ is defined by \hbox{$N(\phi_{\rm CMB},\phi_{\rm
end})\!=\!50$}. Therefore
$P_{\zeta}^{1/2}\!=\!\frac{\sqrt{3}}{4\pi }
\frac{\mu\Lambda^2}{m_p^3}
e^{\left[{100}\left(\frac{m_p}{\mu}\right)^2\right]}$, so that
$\Lambda^2\!=\! \frac{4\pi}{\sqrt{3}} P_{\zeta}^{1/2}
\frac{m_p^3}{\mu}
e^{-\left[{100}\left(\frac{m_p}{\mu}\right)^2\right]}$. Putting
numbers, and restoring explicitly the dependence on
$v''(\phi_{\rm max})$ we obtain
\begin{equation}
\Lambda \simeq 6.5\times 10^{16} GeV  \left(\frac{|v''(\phi_{\rm
max})|}{2}\right)^{1/4} e^{-25|v''(\phi_{\rm max})|}.
\end{equation}
So for consistency with our estimates in
eqs.(\ref{lamI}),(\ref{lamHW}) we need that $|v''(\phi_{\rm
max})|\sim O(1/25)$, a little stronger than the condition that
guarantees a sufficient amount of inflation which we have
encountered previously. This is not a technically unnatural fine
tuning since the gauge coupling at this point is of the same
order.

The index of scalar perturbations is given by
$n_s=1-4\epsilon_{\rm CMB}+2\eta_{CMB} $ and the tensor to scalar
ratio is given by $ r=13.7 \epsilon_{\rm CMB}$. For our model
$\epsilon_{CMB}$ is extremely small, and $\eta_{CMB}\simeq
v''(\phi_{\rm max})$. Consequently,
\begin{eqnarray}
n_S&=& .92 - .08(25 |v''(\phi_{max})|-1) \nonumber \\
r&\simeq&0.
 \label{spectrum}
\end{eqnarray}
Taking as a reasonable range $1/3 \simlt 25|v''(\phi_{max})|
\simlt 3$ we estimate $.76\le n_S\le .97$. This is within the
range of allowed models based on current CMB anisotropy data and
can be verified or falsified  by more accurate data.

Detailed and accurate calculations of the non-perturbative
potential in the central region are not yet available and have to
await improvements in calculation techniques in string theory. In
this context measurements of CMB anisotropies can be a useful
guide to theory.


It is not likely that simple ``large field" inflationary models
in the central region are viable, since the  central region cannot
accommodate large motions of the inflaton. For example
\cite{kinney1}, if the potential is $(\phi/m_p)^p$, the inflaton
field $\phi$ needs to move at least $\Delta(\phi^2)=m_p^2 100 p$.
Assuming  that $\Delta\phi\sim\phi$, $\Delta \phi \sim 10
\sqrt{p}\ m_p $. This is already too large for $p=2$, since this
requires a central region size of at least about $30 m_p$, which
is unlikely. If CMB anisotropy data selects large field models
(currently disfavored by the data \cite{kinney2}) then all simple
models of central region inflation will be ruled out.


This research is supported by grant 1999071 from the United
States-Israel Binational Science Foundation (BSF), Jerusalem,
Israel.  SdA is supported in part by the United States Department
of Energy under grant DE-FG02-91-ER-40672. EN is supported in
part by a grant from the  budgeting and planning committee of the
Israeli council for higher education.


\begin{thebibliography}{99}

\bibitem{BG}
P.~Binetruy and M.~K.~Gaillard,
Phys.\ Rev.\ D {\bf 34}, 3069 (1986).


\bibitem{BS}
R.~Brustein and P.~J.~Steinhardt,
Phys.\ Lett.\ B {\bf 302}, 196 (1993).

\bibitem{BBSMS}
T.~Banks, M.~Berkooz, S.~H.~Shenker, G.~W.~Moore and
P.~J.~Steinhardt,
Phys.\ Rev.\ D {\bf 52}, 3548 (1995).

\bibitem{PBB}
G.~Veneziano,
hep-th/9902097.


\bibitem{MVDV}
A.~Melchiorri, F.~Vernizzi, R.~Durrer and G.~Veneziano,
Phys.\ Rev.\ Lett.\  {\bf 83}, 4464 (1999).

\bibitem{bda2}
R.~Brustein and S.~P.~de Alwis,
Phys.\ Rev.\ Lett.\  {\bf 87}, 231601 (2001).

\bibitem{bda1}
R.~Brustein and S.~P.~de Alwis,
Phys.\ Rev.\ D {\bf 64}, 046004 (2001).


\bibitem{vilenkin}
A.~Vilenkin,
Phys.\ Rev.\ Lett.\  {\bf 72}, 3137 (1994).

\bibitem{bdn}
R.~Brustein, S.~P.~de Alwis and E.~Novak, in preparation.

\bibitem{banks}
T.~Banks,
hep-th/9906126.

\bibitem{HM}
J.~A.~Harvey and G.~W.~Moore,
hep-th/9907026.


\bibitem{Moore}
G.~W.~Moore, G.~Peradze and N.~Saulina,
Nucl.\ Phys.\ B {\bf 607}, 117 (2001).


\bibitem{Ovrut}
E.~Lima, B.~Ovrut, J.~Park and R.~Reinbacher,
Nucl.\ Phys.\ B {\bf 614}, 117 (2001).

\bibitem{natural}
F.~C.~Adams, J.~R.~Bond, K.~Freese, J.~A.~Frieman and
A.~V.~Olinto,
Phys.\ Rev.\ D {\bf 47}, 426 (1993).


\bibitem{kolb}
J.~E.~Lidsey, A.~R.~Liddle, E.~W.~Kolb, E.~J.~Copeland, T.~Barreiro and M.~Abney,
Rev.\ Mod.\ Phys.\  {\bf 69}, 373 (1997).

\bibitem{kinney1}
S.~Dodelson, W.~H.~Kinney and E.~W.~Kolb,
Phys.\ Rev.\ D {\bf 56}, 3207 (1997).


\bibitem{kinney2}
W.~H.~Kinney, A.~Melchiorri and A.~Riotto,
Phys.\ Rev.\ D {\bf 63}, 023505 (2001).






\end{thebibliography}
\end{document}